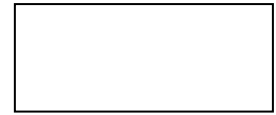

# A method to identify Critical Resources: illustration by an industrial case

*BARBARA LYONNET\*,* Laboratoire SYMME Université de Savoie, Annecy le Vieux, France
Barbara.Lyonnet@univ-savoie.fr
*MAURICE PILLET,* Laboratoire SYMME Université de Savoie, Annecy le Vieux, France
Maurice.Pillet@univ-savoie.fr
*MAGALI PRALUS,* Laboratoire SYMME Université de Savoie, Annecy le Vieux, France
Magali.Pralus@univ-savoie.fr
*LUDOVIC GUIZZI,* Laboratoire IREGE, Université de Savoie, Annecy le Vieux, France
Ludovic.Guizzi@aed.cg74.fr
*GEORGES HABCHI,* Laboratoire SYMME Université de Savoie, Annecy le Vieux, France
Georges.Habchi@univ-savoie.fr

The session type selected for the paper presentation:

**1) "Dialogue" Session**                                        ☐

**2) "Author's Presentation" Session**               ☒

# A METHOD TO IDENTIFY CRITICAL RESOURCES: ILLUSTRATION BY AN INDUSTRIAL CASE

*BARBARA LYONNET\*,* Laboratoire SYMME Université de Savoie, Annecy le Vieux, France
Barbara.Lyonnet@univ-savoie.fr
*MAURICE PILLET,* Laboratoire SYMME Université de Savoie, Annecy le Vieux, France
Maurice.Pillet@univ-savoie.fr
*MAGALI PRALUS,* Laboratoire SYMME Université de Savoie, Annecy le Vieux, France
Magali.Pralus@univ-savoie.fr
*LUDOVIC GUIZZI,* Laboratoire IREGE, Université de Savoie, Annecy le Vieux, France
Ludovic.Guizzi@aed.cg74.fr
*GEORGES HABCHI,* Laboratoire SYMME Université de Savoie, Annecy le Vieux, France
Georges.Habchi@univ-savoie.fr

*Abstract-* **The aim of this study is to develop a method that would enable the company to prioritize the means contributing the most to its performance. The proposed method is based on the profit margin (an economical performance measure of the company), the customer's risk, the costs of maintenance and the employee's safety. The prioritization method of resources was applied to the data obtained from a small subcontracting business in mechanics. The theoretical foundations of this method are based on a multi-criteria approach using the attribution of criticality indexes for nine criteria linked to the financial loss.**

*Keywords: multi-criteria approaches, critical failure factors, manufacturing systems, production management*

## I. INTRODUCTION

Nowadays, one of the essential concerns for a company is the improvement of its economic performance and one of the pillars of this performance is the resources of the company's manufacturing system. The ability to control these resources constitutes a success key for the company's competitiveness. Therefore, to ensure the required availability for production and to meet the customer's requirements, identifying the critical resources appears to be a crucial task. For example, the first aim of an airline is to make its planes fly. In this case, it seems easy to identify the critical resource. However, in the case of a small business with several means of production and many different products and customers, the prioritization of the resources is less obvious. The major interest in identifying the critical resources lies in helping managers to focus on the efficient actions and on the problems that penalize the global performances of the company. Significantly, the company has to constantly review its production system, to be flexible and able to apply quick and right decisions. Giving the lack of time and the more and more competitive market, it becomes necessary to know the resources that cause the greatest loss of profit faced to disruptions.

Different studies have shown a particular interest in the prioritization of the physical resources of companies [1], [2], [3], [4]. One of them suggests prioritizing the physical resources of an agribusiness company's according to the PIEU method, developed by Lavina [4], [3]. This method enables the classification of a set of equipments by attributing to them the following four criticality indexes: the failure index (P), the importance of the equipment (I), the condition of the equipment (E) and the using rate (U). On the other hand, Chelbi and Ait-Kadi's [1] suggest identifying the criteria to prioritize the resources by the means of an organization method developed by Roy [5]. This method is organised in 4 stages: (1) identifying the set of equipments to be classified, (2) establishing a coherent list of priority criteria, (3) evaluating the performance for each part of equipment according to their global performance, and (4) applying an aggregation procedure to class the equipment according to their global performance. On the basis of this step, 9 prioritization criteria have been identified, such as the contribution of the resource to the flow process, the average of the resource's repair time and the importance of the line, in which the equipment is part [2]. A more recent study, carried out at a production unit of plastic products classifies the equipment according to a multi-criteria matrix weighted coefficients for each part of equipment [1]. The retained criteria are importance of the machine, security and consumption. These studies lead to different ways of prioritization of the resources in the context of the maintenance. Nevertheless, the resources of the company can not be perceived as being only physical, but more accurately as a combination of physical and human ones. Yet, it is the control of the company's human resources, which is at the heart of the competitive advantage [6], [7], [8]. The originality of our method consists to consider directly both human resources and the economical aspect. Indeed, in order to maximise its performance, the company should identify the resources that most influence its economic performance.

Based on a multi-criteria approach, this article suggests a new method of resources prioritization more adapted to economic demands and strategic needs of a company. In order to achieve this objective, the principal criteria of the resources' prioritization will be identified, as linked directly to their impact on the profit loss, taking into account the human resources. The aim of this method is to help the planning of preventive and improvement actions. This method was applied to actual data gathered at a screw cutting company.

## II. PROPOSED METHOD

*A. Resources' prioritization method based on a criticality matrix*

On the basis of a multi-criteria approach founded on a criticality matrix, a new resources' prioritization method is developed. This method uses a desirability function which makes it easy and quick. This is very important since the company undergoes constant changes and then requires a perpetual knowledge of resources which impact the economical performance. Moreover, the using of a criticality matrix is perfectly adapted to a multi-criteria approach.

To build the criticality matrix, we explored the principal criteria which are necessary for the application of the method. At first, we propose to identify the criteria directly impacting the company's profit margin. Then, we define additional qualitative criteria necessary to improve this matrix.

The proposed method allows to quantify the loss of profit margin caused by the different types of stop for each one of the company's resources. As the loss of the company's profit margin depends on a complex mix between the production loss and the combination of multiple factors, the set of the criteria directly influencing the margin will have to be looked for.

Hence, the profit margin (PM) can be calculated in the following way:

$$PM = \sum_{i=1}^{n} Pqi.RPi.pPMi$$

Where,

*n:* number of products types produced by the company
*Pqi*: estimated produced quantity of the product i ($Pqi = TPqi - LPi$)
*TPqi*: theoretical production of the product i
*LPi*: estimated loss of production of the product i, linked to the physical and human failures
*SPi*: Selling price of product i
*pPMi*: percentage of the profit margin for product i

Therefore, in order to identify the critical resources of a company, the following five criteria are to be taken into consideration: the selling price, the percentage of profit margin, the reliability rate, the proportion of the quantity of products manufactured by machine and the unavailability. However, these five criteria are not the only ones which can used to measure the risk of a financial loss for a company. Indeed, the failures of some resources can generate a risk in customer's satisfaction and then a loss of market. Some resources present a risk to the safety of the employees of company. Therefore, the following four other important criteria are added to the first five already cited: the customer's risk, the safety of employees, the uniqueness of production means and the costs of maintenance.

Thus, the detailed presentation of the nine identified criteria is as follows:

1) The selling price (SP),

In order to build the criticality matrix the company has to identify the selling price of each product type.

2) The percentage of profit margin (pPM),

This criterion corresponds to the estimated percentage of the profit margin for each product type. It is evaluated by the company according to the estimated cost systems.

3) The estimated reliability rate (Err):

The estimated reliability rate is calculated in the following way:

$$\text{Err} = \frac{TPqi.Ct}{Ot}$$

Where:　*TPqi*: theoretical production of product i
　　　　*Ct*: cycle time (the processing time for product i)
　　　　*Ot*: opening time

4) The proportion of the quantity of products that can be manufactured (pM)

This criterion is calculated in the following way:

$$pMj = \frac{TPqj}{\sum_{i=1}^{n} TPqi}$$

Where-　n: number of machine
　　　　*pMj*: proportion of parts for machine j.
　　　　*TPqi*: theoretical production of machine i

5) Unavailability (material and human failures)

The loss of production linked to the failures depends on the machine reliability rate. Indeed, it is possible that the failures do not have any influence on the produced quantity when the reliability rate of the resource is low.

The availability criterion, which is being referred to in this study, takes into account the risk linked to the absence of human competences. A new idea suggested in this study is to calculate the human-machine Mean Time Between Failures (MTBF).

In order to consider the risk of the production loss linked to both physical and human resources, the theory of reliability science is applied here.

The human resource is considered in the MTBF calculation only if it demonstrates that it is the unique competence for a given machine. Indeed, in this case if the human resource having the specific skills for a physical resource is absent, the resource to which it is habitually assigned is then stopped. On the other hand, where several human resources have the necessary skills for a given machine, the risk linked to the absence of a competent human resource is then without consequence. The main two parameters needed to evaluate the reliability function are: the repair rate ($\mu$) and the failure rate ($\lambda$).

If we consider an exponential distribution, the mathematical expectation E(t) between failures, which represents the MTBF, is: $MTBF = E(t) = \frac{1}{\lambda}$

And the mathematical expectation for downtimes E(tar), representing the Mean Time To Repair (MTTR) is:

$$MTTR = E(tar) = \frac{1}{\mu}$$

The risk of downtimes for the machines having a unique human competence is represented by the human-machine MTBF and MTTR ($MTBF_{HM}$, $MTTR_{HM}$).

Both of the elements – Machine and Human $M_{HM}$ – are represented by a serial system from the viewpoint of reliability, consequently:

$\lambda_{HM} = \lambda_H + \lambda_M$

$\lambda = \frac{\text{Numbers of downtimes}}{\text{Observed time}}$

With:
Number of human downtimes = number of absences
Number of machine downtimes = number of failures
Then,

$MTBF_{HM} = \frac{1}{\lambda_M + \lambda_H}$

For the evaluation of the $MTTR_{HM}$ the average of the weighted downtime is given by:

$MTTR_{HM} = \frac{\lambda_H \times MTTR_H}{\lambda_H + \lambda_M} + \frac{\lambda_M \times MTTR_M}{\lambda_H + \lambda_M}$

For the evaluation of the unavailability, the calculation is given by:

$I = 1 - \frac{MTBF}{MTBF + MTTR}$

In some conditions, breakdowns do not have any effect on the produced quantity. Consequently, the related criteria influencing the profit margin are only considered when the availability ratio of the resource is lower than the estimated reliability rate.

6) Customer's risk

This criterion can be related to the strategy adopted to satisfy a specific customer or the customers representing a majority of the company's market. The aim of this criterion is to identify the resources which harm to the satisfaction of most essential customers.

7) Safety of employees

It is necessary to follow resources which could put in jeopardy the employees of a company. In the extreme case, if not considered, the employee's safety could generate a financial loss and harm the survival of the company. Failures of some resources are more dangerous than others.

8) Uniqueness of production means

Several resources could have a specific and single manufacturing process. If this kind of resource is stopped some products cannot be produced, which could harm the satisfaction of customers involving in some cases a loss of market.

9) Costs of maintenance

The costs assigned to the maintenance are different according to the technology or the age of a machine; some resources can generate a higher financial loss. This criterion is necessary to build a hierarchical organisation of the different machines of a company.

This method of resources' prioritization – elaborated with the help of a criticality matrix – is based on the attribution of criticality indexes to these nine criteria. These criticality indexes are attributed thanks to a desirability function (d). For the evaluation of the desirability $d_i$ the calculation is given by:

$$d_i = \frac{v_j}{\max(v_j)}$$

Where,
$V_j$: the value of criterion j

The highest value of which is rated as 1. The overall desirability D, another value between 0 and 1, is defined by combining the individual desirability values. The overall desirability is defined by the geometric mean:

$$DG = (d_1 . d_2 . d_3 ... d_k)^{1/k} = \left(\prod_{i=1}^{k} d_i\right)^{1/k}$$

This overall desirability corresponds to the criticality of the studied resources.

## III. APPLICATION AND RESULTS

*A. Application of the proposed criticality method to the data of a screw cutting company:*

The example of a screw cutting company has been chosen. This company is located at the heart of the Arve valley (Haute-Savoie region, France) and then constitutes a particularly interesting application field. The Arve valley is considered to be one of the principal local French productive systems. The companies of the valley generate more than 60% of the French turnover of the screw cutting activity, i.e. fabrication of machine parts out of essentially metal materials. In the case of most of the parts manufactured in a screw cutting company, the product generally undergoes the following two successive transformation operations: screw cutting and washing. One of the particularities of the screw cutting company resides in the configuration of its production system (cf. Figure 1).

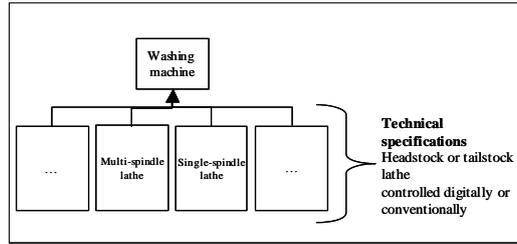

**Figure 1**. Configuration of the producing department of the examined screw cutting company.

| Resource/Machine(M) | Product(P) | Selling price(in euros) | Cycle time(in minutes) | Average percentage of the margin by product | MTBF Human-Machine (in minutes) | Estimated Reliability rate | Production carried out (in number of parts) | Proportion/contribution to the total production |
|---|---|---|---|---|---|---|---|---|
| M1 | P1 | 0.07 | 0.06 | 10% | 4500 | 0.84 | 2784 | 41% |
| M2 | P2 | 0.9 | 0.75 | 15% | 4500 | 0.93 | 249 | 4% |
| M3 | P3 | 0.23 | 0.07 | 2% | 4160 | 0.73 | 2112 | 31% |
| M4 | P4 | 0.5 | 0.11 | 5.50% | 4500 | 0.95 | 1728 | 25% |
| M5 | All | 0.425 | 0.017 | 8% | 4500 | 0.58 | 6873 | 100% |

**Table 1.** Data gathered in the studied company

| Machine (M)/Resource | Opening time in minutes per year | Average number of failures per year | Average down time in minutes | MTBF Machine in minutes | MTBF Human-Machine in minutes | Availability |
|---|---|---|---|---|---|---|
| M1 |  | 45 |  | 4500 | NA | 0.65 |
| M2 |  | 45 |  | 4500 | NA | 0.65 |
| M3 | 202500 | NA | 2400 | NA | 4160 | 0.63 |
| M4 |  | 45 |  | 4500 | NA | 0.65 |
| M5 |  | 45 |  | 4500 | NA | 0.65 |

**Table 2.** Failure parameters

| Machine (M)/resource | Profit margin | | | | | Customers | | Employees | Maintenance | Criticality Overall desirability (OD) |
| | (If availability< reliability rate) Unavailability | Flow-process grid contribution | Estimated reliability rate | Selling price | Previsional percentage of the profit margin | Uniqueness of production's mean | Consequence customer | safety | Costs of maintenance | |
|---|---|---|---|---|---|---|---|---|---|---|
| M5 | NA | NA | NA | NA | NA | 1 | 1 | NA | 1 | 1.00 |
| M4 | 0.97 | 0.25 | 1 | 0.56 | 0.37 | NA | NA | NA | NA | 0.55 |
| M2 | 0.97 | 0.04 | 0.98 | 1 | 1 | NA | NA | NA | NA | 0.52 |
| M1 | 0.97 | 0.5 | 0.88 | 0.08 | 0.67 | NA | NA | NA | NA | 0.47 |
| M3 | 1 | 0.3 | 0.77 | 0.26 | 0.13 | NA | NA | 1 | NA | 0.45 |

NA: No Applicable

**Table 3.** Classification of the resources according to the proposed method

In this company, three types of resources are studied: two multi-spindle lathes, two single-spindle lathes and a washing machine. As it is depicted by the diagram 1, the lathes are independent from each other. Another particularity of this system is the importance of the only resource contributing to the flow-process of the company, i.e. the washing machine.

*B. The company's data:*

The data presented below are those gathered during the observations carried out in this company and the interviews with employees (table 1). The data used to calculate the unavailability of machines are summarized in table 2. The examined company processes its production during a period of 45 weeks per year. The average maximal number of failures occurring per month is 4, i.e. 45 per year. The opening time during the period of 45 weeks comprises 202500 minutes. The maximal downtime of a machine is estimated to be 40 hours (2400 minutes).

The failure rate of each machine has been calculated on the basis of the maximal number of stops – rather than the average number – in order to ensure the examination of the maximal impact of a failure on the production. Likewise, the repair rate has been calculated on the basis of the maximal time of the repair.

Except for machine M3, the machine MTBF is equal to 4500 minutes. For the machine M3 a human-machine MTBF has been calculated.

More, this new approach is carried out according to the other following data:
- Average number of absences per person and per year: 1.97
- Average duration of an absence per year: 5 days
- Time worked by a person per year: 108 000 minutes (8h x 5days x 45weeks x 60)
- Human failure rate: $\lambda H = \frac{1.96}{108000} = 1.83 \times 10^{-5}$
- Machine failure rate: $\lambda H = \frac{45}{202500} = 2.22 \times 10^{-4}$
- Human-machine MTBF:
$$MTBFHM = \frac{1}{\lambda M + \lambda H} = \frac{1}{2,40 \times 10^{-5}} = 4160 mn$$

- Human-machine MTTR: 2400 minutes

*C. Criticality matrix: obtained results*

A weighted scale is used; the highest value is rated as 1. To evaluate the risk linked to the loss of profit margin, the resource M5 which has an estimated rate lower than the availability is not considered. The risk of the production loss of this machine linked to the risk of failures calculated by the MTBF and the MTTR is null. For the other resources the desirability is calculated (cf. Table 3). For example, for the criterion related to the selling price, the calculation of the scale is as follows:

The highest selling price is that of the product produced by the machine M2 (0.9 €), and consequently, its rating value is the highest (1).
In the case of the machine M4, whose product selling price is 0.5 €, the rating is carried out through desirability (d): $dM4 = \frac{0.5}{0.9} = 0.56$

The final index of the criticality is calculated for each machine according to the overall desirability for nine prioritization criteria that have been retained. It enables the prioritization of the resources in relation to their impact on the company's profit margin and on the financial loss.
The identified resource as critical is the machine M5. This is a uniqueness of production means and it contributes to the satisfaction of all customers of the company since it is essential to the production of all products.

## IV. DISCUSSION

*A. Criticality matrix: nine criteria*

Our new prioritization method is based on the attribution of the criticality index for the 5 quantitative criteria directly influencing the company's profit margin and 4 qualitative criteria. We tested a part of our method based on the quantitative criteria influencing the profit margin by simulation. Prioritization results obtained by simulation are the same of those obtained with our prioritization method. In the present case, the resources generating the biggest financial loss are resources M4 and M5. In order to combine quantitative and qualitative criteria we used the desirability approach by the calculation of an overall desirability. The overall desirability allows combining different kinds of data. This approach transforms estimated data in the same scale between 0 and 1. Comparing overall desirability of each resource is then easier.

*B. Qualitative criteria*

The qualitative criteria used in our method are targeted on the risk of unsatisfied customers, uniqueness of production means, costs of maintenance and the risk linked to employees' safety. Considering these criteria seems essential since it generates a risk of financial loss for the company. Indeed, in some cases, the no satisfaction of a customer can lead to generate a financial loss linked to the break of a contract. The criterion linked to the uniqueness of production means is necessary in the methods of resources' prioritization [1]. In the present case the resource failure which have a uniqueness of production means generates a no satisfaction of all customers; it is necessary to ensure the availability of this resource. The third criterion such as the risk linked to the employees' safety is essential in the methods of resource's follow-up. This criterion can, in the extreme case, generate a financial loss. Moreover the taking into consideration of the employees' safety contributes directly to the improvement of work conditions that constitutes a competitive advantage.

*C. Consideration of the risks linked to the human resources:*

One of the criteria used in our method of the resources' prioritisation is the unavailability linked to the failures. It seems advantageous to consider the company's human resources by calculating the human-machine MTBF on the basis of the number of absences of the company's employees. Therefore, our model is believed to be more representative of the company and of all of the resources influencing its profit margin. The human-machine MTBF, calculated for the machine M3 (the only machine presenting a unique human resource) is lower than the machine MTBF of all the other machines. In fact, as it was expected, the consideration of the human "failure" risk increases the number of possible stops. Until now, this notion of human "failure" has not been considered in any classification study, thus neglecting a risk for the company. Obviously, this risk is linked to the issue of absenteeism, so common in every company.

*D. Criticality matrix: improvement actions*

Several preventive actions can be identified in order to improve the performance due to the risks linked to financial loss. In order to improve the availability of its resources, the company would implement preventive actions of maintenance, Maintenance Based on Reliability (MBF) or (RCM) Reliability Centred Maintenance [9], and also with a long-term orientation, Total Productive Maintenance (TPM) [10].

The failure rate of a machine is variable and depends on dysfunctions of production and organisation. This variation depends on different stops of resources as breakdowns, stops for quality control, and machine starvation. The decrease and the disappearance of risks are thus based on the control of maintenance (RCM and TPM), the quality control (SPC) [11] and supply management. The right application of these methods is based on a good knowledge of the level of quality required by the customer. Indeed, the customer could appraise defective products considered being acceptable by the supplier [12]. On the other hand, sometimes it is useless to reduce defects which will not be perceived like such by the customer. According to [13], the companies have difficulties to define the desired value by the customer. Indeed the value is variously defined according to speakers, each one defines the value in its way, according to its needs. It is necessary, to define with customers, criteria which will define the defects and thus the desired value [13]. These actions will allow decreasing the risk linked to customer's loss. The risk linked to the safety of employees could be decreased by the implementation of preventive actions, such as the fast supply of spare parts, a plan of safety for resources identified as critical [14], [15].

## V. CONCLUSION

This study has enabled the introduction of a new method for the prioritization of the companies' critical resources according to their profit margin, their customer satisfaction, the costs of maintenance and the employee's safety. This method rests upon a multi-criteria approach based on the use of a criticality matrix, composed of 9 criteria directly linked to the financial loss. This step is advantageous in its referring to the profit margin itself, an essential parameter of the economical performance. More importantly, it takes into account – by calculating the human-machine MTBF – the impact of the absence of human skills on the material resources.

The desirability function approach is advantageous to combine qualitative and quantitative data. This prioritization method is simple and fast in use, and addresses to the companies wishing to know at any moment which resource generates the biggest loss of the profit margin, as much for the sake of their everyday management, as for the development of a new strategy. Having quickly prioritized its resources, a company can introduce the actions aiming at improving the situation, more centred around the resource itself, in order to increase the economical performance. Our method is adapted to the application case. This company produce in mass. To validate this method in company of different types and environment we will apply it in other companies. On the basis of this prioritization of the resources, a company can ensure the required availability for production in order to increase the economical performance, the customer satisfaction and the safety's employee.